\documentclass{atlasnote} 

\usepackage{atlasphysics} 

\usepackage{hyperref}
\usepackage{xspace}
\usepackage{epsfig}
\usepackage[numbers,merge,sort&compress]{natbib}

\renewcommand{\ifb}{fb$^{-1}$\xspace}
\renewcommand{\pt}{\ensuremath{p_{T}}\xspace}
\newcommand{\tth}{\ensuremath{t\bar t H}\xspace}

\skipbeforetitle{100pt}

\title{\boldmath Analysis of \ttbar$H$ Events at $\sqrt{s}=14$~TeV with $H\rightarrow WW$}

\usepackage{authblk}
\author[a]{P.~Onyisi}
\author[b]{R.~Kehoe}
\author[a]{V.~Rodriguez}
\author[a]{Y.~Ilchenko}

\affil[a]{\footnotesize University of Texas at Austin, Austin, TX, USA}
\affil[b]{\footnotesize Southern Methodist University, Dallas, TX, USA}

\date{}

\abstracttext{

We study the process $pp\rightarrow\ttbar H$ at $\sqrt{s}=14$~TeV where 
$H\rightarrow WW$.  We focus on final states that include three or four
leptons.  Signal and background samples are generated at leading 
order (LO) and normalized to next-to-leading order (NLO) calculations.  We employ
the ``Snowmass'' detector simulation and event reconstruction.  Cuts are 
selected which provide a substantial signal yield while suppressing the
main backgrounds.  Systematic uncertainties are estimated for the 
overall normalization of signal and background processes.  With no improvement in the current theoretical uncertainties on $\sigma(\ttbar H)$, we project
a precision on a top quark Yukawa coupling of approximately $16\%$ in 300~\ifb and
 $9\%$ in 3000~\ifb.  If the theoretical uncertainties are halved, these improve to $14\%$ and $6\%$, respectively.}

\begin{document}


\section{Introduction}
\label{sec:intro}
In the standard model, fundamental fermions are given mass by their
interaction with the Higgs boson.  The Yukawa coupling governing this
interaction is proportional to the mass of the fermion, as
$Y_f = \sqrt{2} m_f/v$ where $v$ is the vacuum expectation value of the
Higgs field.  Indirectly, the Yukawa coupling can be inferred
from the measured mass of the particle, which in the case of the top
quark \cite{Aaltonen:2012ra} 
yields $Y_t = 0.996\pm0.005$ at leading order.  However, this measure is indirect,
and is uncertain due to the nature of the theoretical mass being measured.  
$Y_t$ can be measured via the loop processes $gg \to H$ and 
$H \to \gamma\gamma$, but additional particles (possibly non-Standard Model) 
may contribute to the loops (for example $H \to\gamma\gamma$ is already 
dominated by a $W$ loop), so it is not possible to distinguish deviations in 
the apparent value of $Y_t$ from other new physics.
Therefore, it is important to measure the top quark Yukawa coupling directly 
through a tree-level process,
and the most effective way to do this is through production of top quark
pairs in conjunction with a Higgs boson.  

This paper describes an analysis of such production at the LHC at 
$\sqrt{s} = 14$~TeV with integrated luminosity of 300~fb$^{-1}$
and 3000~fb$^{-1}$.  We have concentrated on final states available
via the decay $H\rightarrow WW$, due to its high branching fraction
and unique leptonic signature.  Specifically, we discuss the potential
for analysis of final states with three and four leptons.  The
theoretical and sample parameters for signal and background processes
are described in Sec.~\ref{sec:SandB}. 
We describe the event selection in Sec.~\ref{sec:EvtSel}.
The estimated statistical precision, a discussion of systematic
uncertainties, and the projected sensitivities of the analysis
are given in Sec.~\ref{sec:Results}.

\section{Signal and Background Processes}
\label{sec:SandB}
\subsection{Monte Carlo Event Samples}
\label{sec:samples}

The signature of Higgs boson production in conjunction with
top quark pairs provides a complex array of diverse final states.
The $H\rightarrow WW$ decay can yield events with multiple isolated leptons in a highly distinctive signature including
two $b-$jets, additional light quark jets, and \met.  The primary
physics backgrounds giving such a signature arise from
\ttbar$Z$ and \ttbar$W$ production.  At a much smaller level, \ttbar$WW$ 
production is also a concern.  For these background processes, we have generated
Monte Carlo samples at $\sqrt{s}=14$ TeV using MadGraph 5 v2 beta \cite{Alwall:2011uj} with Pythia 6 showering \cite{Sjostrand:2006za}, as itemized in 
Table~\ref{tab:samples}.
We similarly generated $\ttbar H$ samples with $H$ decays to $WW$, $ZZ$ 
and $\tau\tau$, all of which can result in multilepton events.  To combine the samples we use 
branching fractions for Higgs decay from the LHC Higgs Cross Section 
Working Group \cite{Dittmaier:2011ti,Djouadi:1997yw,Bredenstein:2006rh,Actis:2008ts,Denner:2011mq}.  The $H \to \tau\tau$ and $H\to ZZ$ decays contribute non-negligibly to the observable signal.

\begin{table}
\begin{center}
\caption{\label{tab:samples}Monte Carlo samples, beyond the official Snowmass background samples, used for signal and background estimation.  All
samples generated at LO in MadGraph, using CTEQ6L1 PDFs and Pythia 6 for showering.} 

\begin{tabular}{cc}
\hline\hline
{\sc Process} & {\sc Scale}  \\
\hline
\ttbar $H+\le 1p$    & $m_t + m_H/2$\\
\ttbar $W+\le 2p$    & $m_t + m_W/2$  \\
\ttbar $ll+ \le 2p$   & dynamic (geometric mean of $\sqrt{m_i^2+\pt(i)^2}$ for final state particles)  \\
\ttbar $WW$      & $m_t + m_W$  \\
\hline\hline
\end{tabular}
\end{center}
\end{table}

In addition to these samples, we model other backgrounds using
samples from the official Snowmass Energy Frontier generation \cite{Anderson:2013kxz,Avetisyan:2013onh,Avetisyan:2013dta}.  This includes a large sample of
\ttbar$+jets$ events.  The primary instrumental background
comes from this process, where a lepton is either incorrectly
identified from one of the jets or a non-isolated lepton is misconstrued
as isolated.  

All our samples were processed through the Delphes fast detector simulation 
\cite{deFavereau:2013fsa} 
(version 3.0.9) using the official configurations for the ``Snowmass detector'' in the $\langle\mu\rangle = 0$, 50, and 140 pileup scenarios\footnote{Many thanks to Sergei Chekanov (ANL) for handling this step.} \cite{Anderson:2013kxz}.

\subsection{Cross Sections}
\label{sec:xsecs}

For the samples listed in Table~\ref{tab:samples}, we extracted LO cross 
sections from MadGraph as given in Table~\ref{tab:xsecs}.  When possible we
used aMC@NLO \cite{amcatnlo} to obtain NLO cross sections; these are also shown in Table~\ref{tab:xsecs}.  The obtained NLO cross section for $\ttbar H$ production is in good agreement with previous determinations \cite{Dittmaier:2011ti,Reina:2001sf,Beenakker:2001rj,Beenakker:2002nc,Dawson:2003zu}.
For Snowmass Energy Frontier background samples, we use the official LO cross sections from the event generation.  The per-event weights in the samples include $k$-factors to bring the inclusive cross sections to NLO.
\begin{table}
\begin{center}
\caption{LO and NLO cross sections for generated
signal and primary physics backgrounds.} 
\label{tab:xsecs}
\begin{tabular}{cccc}
\hline\hline
{\sc Process} & {\sc LO} $\sigma$ (fb) & {\sc NLO} $\sigma$ (fb) & {\sc k-Factor} \\
\hline
\ttbar $H+1p$    & 533.6   & 609.9    & 1.14 \\
\ttbar $W+2p$    & 548.7   & 706.2    & 1.29 \\
\ttbar $ll+2p$   & 74.5    & 74.1     & 0.99 \\
\ttbar $WW+1p$   & 10.4    & n/a      & ---  \\
\hline\hline
\end{tabular}
\end{center}
\end{table}

\begin{figure}
    \centering
\includegraphics[width=0.5\linewidth]{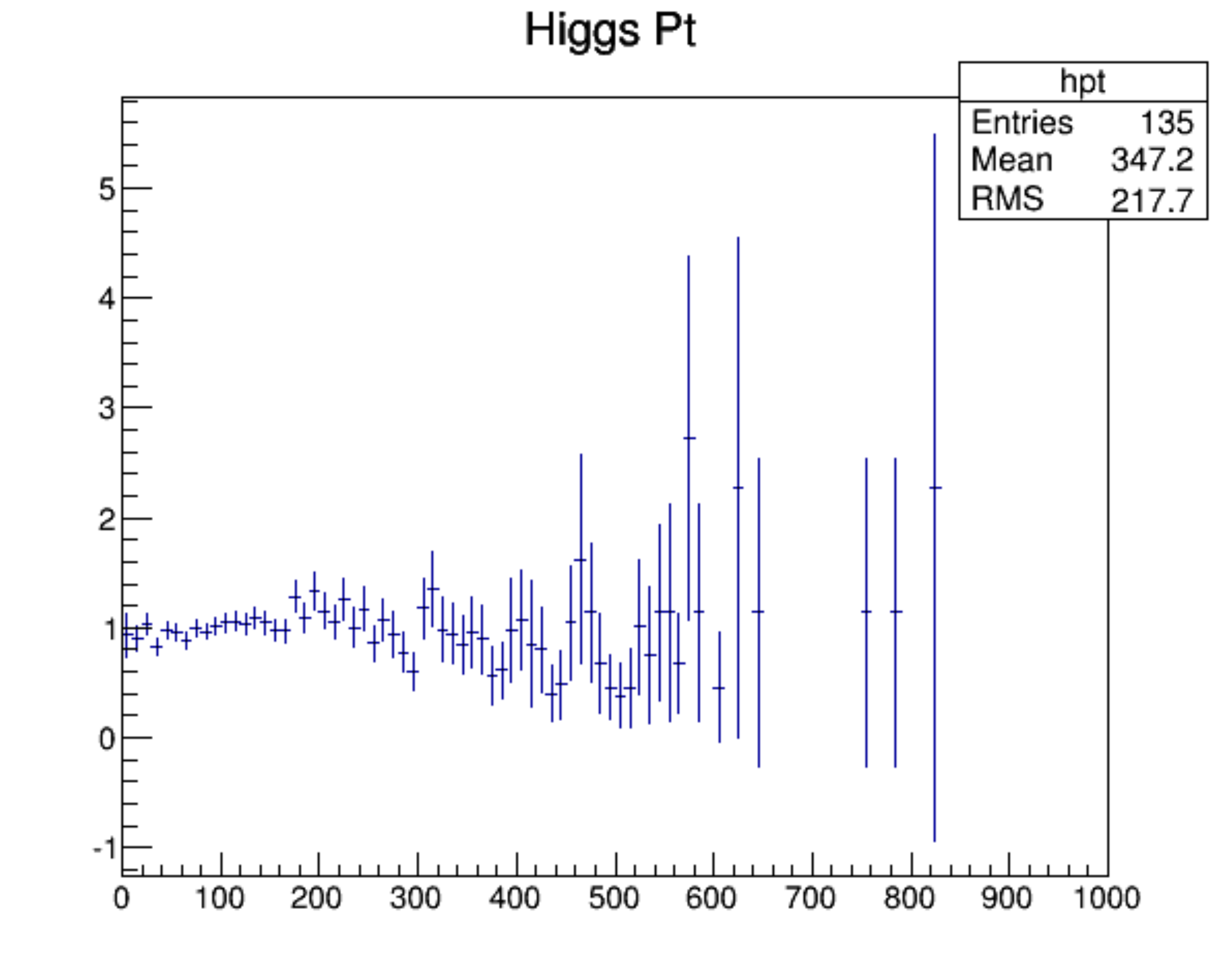}%
\includegraphics[width=0.5\linewidth]{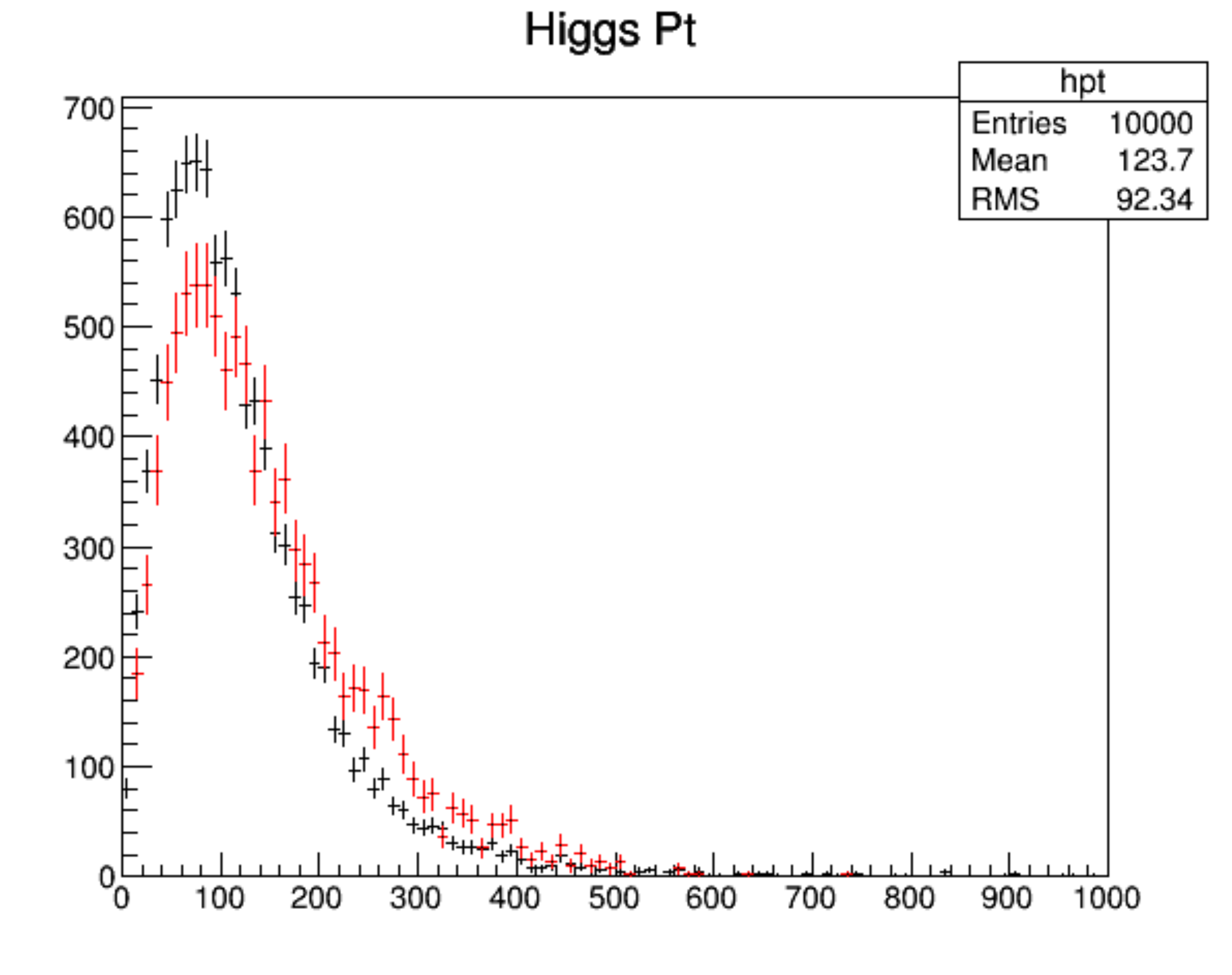}
    \caption{The ratio of NLO to LO for the Higgs boson $p_T$ distribution (left).
     The $p_T$ distribution of Higgs bosons (black) and Z bosons (red)
     in signal and background samples, respectively (right).}
\label{fig:HvZ}
\end{figure}

We have examined the kinematics of the top quarks and associated massive bosons
for the LO and NLO cases, particularly for \ttbar$H$ and \ttbar$\ell\ell$ cases
where the latter is dominated by \ttbar$Z$ production.  We observe that
the differential distributions are not significantly altered by the absence
of the higher order terms in the LO samples, despite the large impact on the
total cross section for signal.  The ratio of NLO to LO vs. the Higgs boson
$p_T$ is shown in Fig.~\ref{fig:HvZ}.
Differences between signal and background
are evident, as shown in Figure~\ref{fig:HvZ} which illustrates the harder
$p_T$ of the Z boson in \ttbar$\ell\ell$ events as compared to that of the
Higgs boson in \ttbar$H$ events.  Partly for this reason, we focus on
developing a strategy for controlling this background.

\section{Event Selection}
 \label{sec:EvtSel}
  
\begin{table}
\begin{center}
\caption{\label{tbl:cuts}Cuts defining each of the signal and control regions of the 300 \ifb analysis.  See the text for detailed descriptions of the cuts.}
 \begin{tabular}{l|cccccccccc}
 \hline\hline
  Cut & \rotatebox{90}{$3\ell$ SR $Z$-depleted} & \rotatebox{90}{$3\ell$ SR $Z$-enriched} & \rotatebox{90}{$3\ell$ SR $Z$-superenriched} & \rotatebox{90}{$3\ell$ $t\bar t Z$ CR} & \rotatebox{90}{$3\ell$ $WZ/ZZ$ CR} & \rotatebox{90}{$3\ell$ high mass} & \rotatebox{90}{$4\ell$ SR $Z$-depleted} & \rotatebox{90}{$4\ell$ SR $Z$-enriched} & \rotatebox{90}{$4\ell$ SR $Z$-superenriched} & \rotatebox{90}{$4\ell$ $ZZ$ CR} \\
  \hline
  \# leptons, $\pt > 10$ GeV & \multicolumn{6}{c|}{3} & \multicolumn{4}{c}{4}\\ \cline{2-11}
  \# leptons, $\pt > 25$ GeV & \multicolumn{6}{c|}{3} & \multicolumn{4}{c}{---}\\ \cline{2-11}
  jet counting threshold (GeV) & \multicolumn{6}{c|}{30} & \multicolumn{4}{c}{35} \\ \cline{2-11}
  \# jets & \multicolumn{6}{c|}{$\ge 4$} & \multicolumn{4}{c}{$\ge 2$} \\ \cline{2-11}
  \# loose $b$-tags & \multicolumn{4}{c|}{$\ge 1$} & 0 & \multicolumn{4}{|c|}{$\ge 1$} & 0 \\ \cline{2-11}
  $Z$ candidate & veto & veto & veto & yes & yes & veto & veto & veto & veto & yes\\ \cline{2-11}
  \# opposite sign same flavor pairs & 0 & 1 & 2 & $\ge 1$ & $\ge 1$ & any & 0 & 1 & $\ge 2$ & $\ge 1$ \\ \cline{2-11}
  $m(ll)_{01}$ (GeV) & $\le 70$ & $\le 70$ & $\le 70$ & --- & --- & $\ge 70$ & --- & --- & --- & ---\\
  \hline\hline
 \end{tabular}
\end{center}
\end{table}

\begin{table}[h]
\begin{center}
\caption{\label{tbl:3000cuts}Cuts defining each of the signal and control regions of the 3000 \ifb analysis.  See the text for detailed descriptions of the cuts.}
 \begin{tabular}{l|cccccccccc}
 \hline\hline
  Cut & \rotatebox{90}{$3\ell$ SR $Z$-depleted} & \rotatebox{90}{$3\ell$ SR $Z$-enriched} & \rotatebox{90}{$3\ell$ SR $Z$-superenriched} & \rotatebox{90}{$3\ell$ $t\bar t Z$ CR} & \rotatebox{90}{$3\ell$ $WZ/ZZ$ CR} & \rotatebox{90}{$3\ell$ high mass} & \rotatebox{90}{$4\ell$ SR $Z$-depleted} & \rotatebox{90}{$4\ell$ SR $Z$-enriched} & \rotatebox{90}{$4\ell$ SR $Z$-superenriched} & \rotatebox{90}{$4\ell$ $ZZ$ CR} \\
  \hline
  \# leptons, $\pt > 10$ GeV & \multicolumn{6}{c|}{3} & \multicolumn{4}{c}{4}\\ \cline{2-11}
  \# leptons, $\pt > 20$ GeV & \multicolumn{6}{c|}{3} & \multicolumn{4}{c}{---}\\ \cline{2-11}
  minimum same sign lepton \pt (GeV) & \multicolumn{6}{c|}{30} & \multicolumn{4}{c}{---}\\ \cline{2-11}
  jet counting threshold (GeV) & \multicolumn{6}{c|}{30} & \multicolumn{4}{c}{35} \\ \cline{2-11}
  jet counting $|\eta|$ acceptance & \multicolumn{10}{c}{$|\eta| < 2.5$}\\ \cline{2-11}
  \# jets & \multicolumn{6}{c|}{$\ge 4$} & \multicolumn{4}{c}{$\ge 2$} \\ \cline{2-11}
  $b$-tag jet \pt cut (GeV) & \multicolumn{10}{c}{30}\\ \cline{2-11}
  \# loose $b$-tags & \multicolumn{4}{c|}{$\ge 1$} & 0 & \multicolumn{4}{|c|}{$\ge 1$} & 0 \\ \cline{2-11}
  $Z$ candidate & veto & veto & veto & yes & yes & veto & veto & veto & veto & yes\\ \cline{2-11}
  \# opposite sign same flavor pairs & 0 & 1 & 2 & $\ge 1$ & $\ge 1$ & any & 0 & 1 & $\ge 2$ & $\ge 1$ \\ \cline{2-11}
  $m(ll)_{01}$ (GeV) & $\le 70$ & $\le 70$ & $\le 70$ & --- & --- & $\ge 70$ & --- & --- & --- & ---\\
  \hline\hline
 \end{tabular}
\end{center}
\end{table}

We use the standard Snowmass Delphes reconstructed objects, including the lepton isolation and pileup corrections for jets.  
Leptons are required to have reconstructed $\pt > 10$ GeV and to pass the default isolation requirement.  Electrons must have $|\eta| < 2.47$ and muons $|\eta| < 2.5$.  Opposite-sign lepton pairs with $m(\ell\ell) < 10$ GeV are assumed to arise from meson decays and are removed from the sample.

The different pileup conditions for the 300 \ifb analysis (assumed mean number of interactions $\langle\mu\rangle = 50$) and the 3000 \ifb analysis ($\langle\mu\rangle = 140$) led to different working points being chosen for the cuts.  

Jets are first preselected requiring $\pt > 20$ GeV and $|\eta| < 4.5$.  Any jet that matches a lepton in direction with $\Delta R \equiv \sqrt{\Delta \eta^2 + \Delta \phi^2} < 0.02$ is assumed to actually arise from that lepton and is removed from consideration.  We use the loose $b$-tagging working point to identify $b$ jets: this plateaus at 75\% (69\%) efficiency for $|\eta| < 1.2$ ($1.2 < |\eta| < 2.5$).  After the jet preselection, any lepton closer than $\Delta R = 0.4$ to a jet is considered non-isolated and not considered further.

Preselected events are sorted into various signal and control regions as identified in Tables~\ref{tbl:cuts} and \ref{tbl:3000cuts} for the 300 \ifb and 3000 \ifb analyses, respectively.  Events are sorted into the $3\ell$ and $4\ell$ analyses based on the number of preselected leptons ($\pt > 10$ GeV after the overlap removal described above).  To reduce the background in the $3\ell$ analysis from $t\bar t$ events with an additional ``fake'' lepton, we raise the minimum lepton \pt cut to 25 GeV for the $3\ell$ channel in the 300 \ifb analysis.  For the 3000 \ifb $3\ell$ analysis, the $t\bar t$ background is yet larger due to higher pileup.  In this case, we require that the two leptons with same sign both have $\pt > 30$ GeV, while the lepton of opposite sign to the other two must have $\pt > 20$ GeV.

We expect the signal to give higher jet multiplicity than several of the 
backgrounds ($WZ$, $ZZ$, $t\bar t$, and $t \bar t W$).  Cutting on jet multiplicity is therefore a useful way to separate signal from background.  Pileup adds extra jets to the event and makes jet counting less reliable.  To reduce the impact of pileup we raise the \pt threshold for jets to be counted to 30 (35) GeV for the $3\ell$ ($4\ell$) analysis.  For the 300 \ifb analysis jets in the full preselected pseudorapidity range are counted; for the 3000 \ifb analysis we only consider those with $|\eta| < 2.5$.  These cuts are motivated by the vastly increased purity of the selected jets with respect to generator-level jets from the hard scattering.  For the 300 \ifb analysis, $b$-jet counting is done with the full set of preselected jets (with \pt down to 20 GeV); for the 3000 \ifb analysis only jets with $\pt > 30$ GeV are considered as $b$-tag candidates.

We find $Z$ boson candidates in events by looking for opposite sign, same flavor lepton pairs satisfying 81 GeV $< m(\ell\ell) <$ 101 GeV.  For signal regions we veto events with a $Z$ candidate, while for control regions we require a $Z$ candidate.

Even after the $Z$ veto, there is some contamination from virtual photon and off-shell $Z$ events $\gamma^*/Z \to \ell^+\ell^-$.  We reduce the contamination by separating signal region events into ``$Z$-depleted'', ``$Z$-enriched'' and ``$Z$-superenriched'' samples, where there are respectively 0, 1, and $\ge 2$ pairs of opposite sign, same flavor leptons.  For example, in the $3\ell$ analysis, a $\mu^+ \mu^+ e^-$ candidate would be $Z$-depleted, a $\mu^+ e^+ e^-$ candidate would be $Z$-enriched, and an $e^+ e^+ e^-$ candidate would be $Z$-superenriched.

For the $3\ell$ analysis we exploit the low mass of the Higgs boson and the spin corrrelation of the $W$ bosons produced in $H \to W^+ W^-$.  These two conditions lead the charged lepton pair from a $H \to \ell \nu \ell\nu$ decay to have low invariant mass, typically below 70 GeV; this is insensitive to Higgs boson kinematics.  We choose the opposite sign pair with smallest $\Delta R$ separation and hypothesize that these are Higgs decay daughters; we denote the invariant mass of the pair as $m(\ell\ell)_{01}$.  Our main signal regions require $m(\ell\ell)_{01} < 70$ GeV, i.e.\ they are primarily sensitive to $H \to \ell\nu\ell\nu$ and lepton+jets top pair decay.  To recover some sensitivity to the $H \to \ell\nu j j$ case, we also include a ``high mass'' bin which requires $m(\ell\ell)_{01} > 70$ GeV.  This bin also partially serves to constrain the $t\bar t W$ background.  In the $4\ell$ case, there are more combinatorics (as there are four same flavor opposite sign pairs) and this variable is much less useful, so we omit it.

We add two control regions to normalize important background sources: $t\bar t Z$ 
and $WZ/ZZ$ production.  For this analysis, we make the simplifying assumption that 
all diboson backgrounds can be normalized together.  Both the control regions invert the $Z$ veto.  The $t\bar t Z$ control region is derived from the $3\ell$ analysis, and the $ZZ$ CR from the $4\ell$ analysis.  The lepton, jet, and $b$-tag selections are identical with the corresponding $3\ell$ and $4\ell$ signal region cuts, which would reduce systematic uncertainties in the background normalization associated with lepton efficiencies, jet reconstruction and energy scale, and $b$-tagging efficiency and mistag rate.

Plots of some variables in the 300 \ifb analysis are shown in Figures~\ref{fig:3lplots} and \ref{fig:4lplots}.  The ultimate yields in each signal and control region are shown in Figures~\ref{fig:yields} (300 \ifb) and \ref{fig:3000yields} (3000 \ifb).

\begin{figure}
\begin{center}
\includegraphics[width=.5\linewidth]{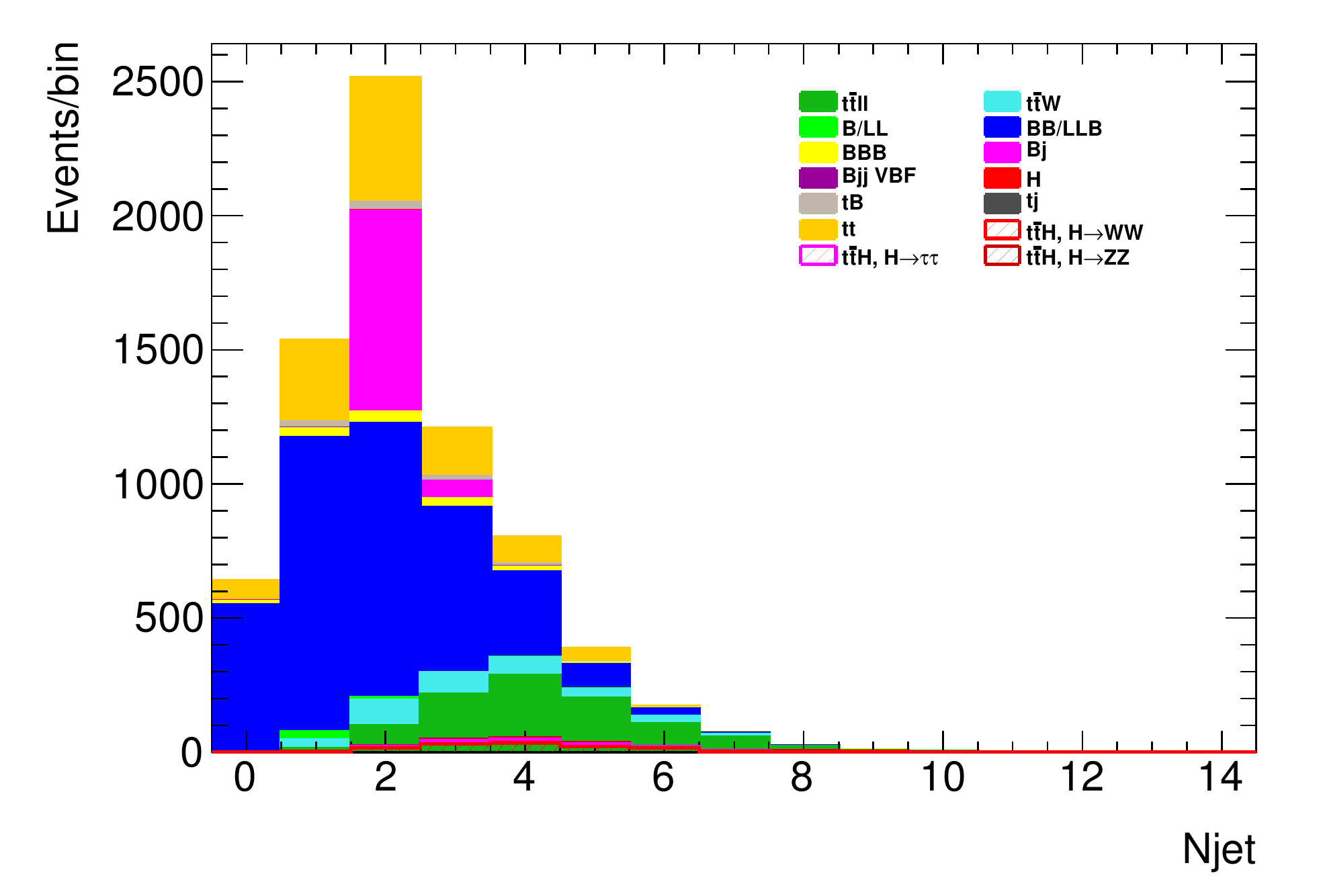}%
\includegraphics[width=.5\linewidth]{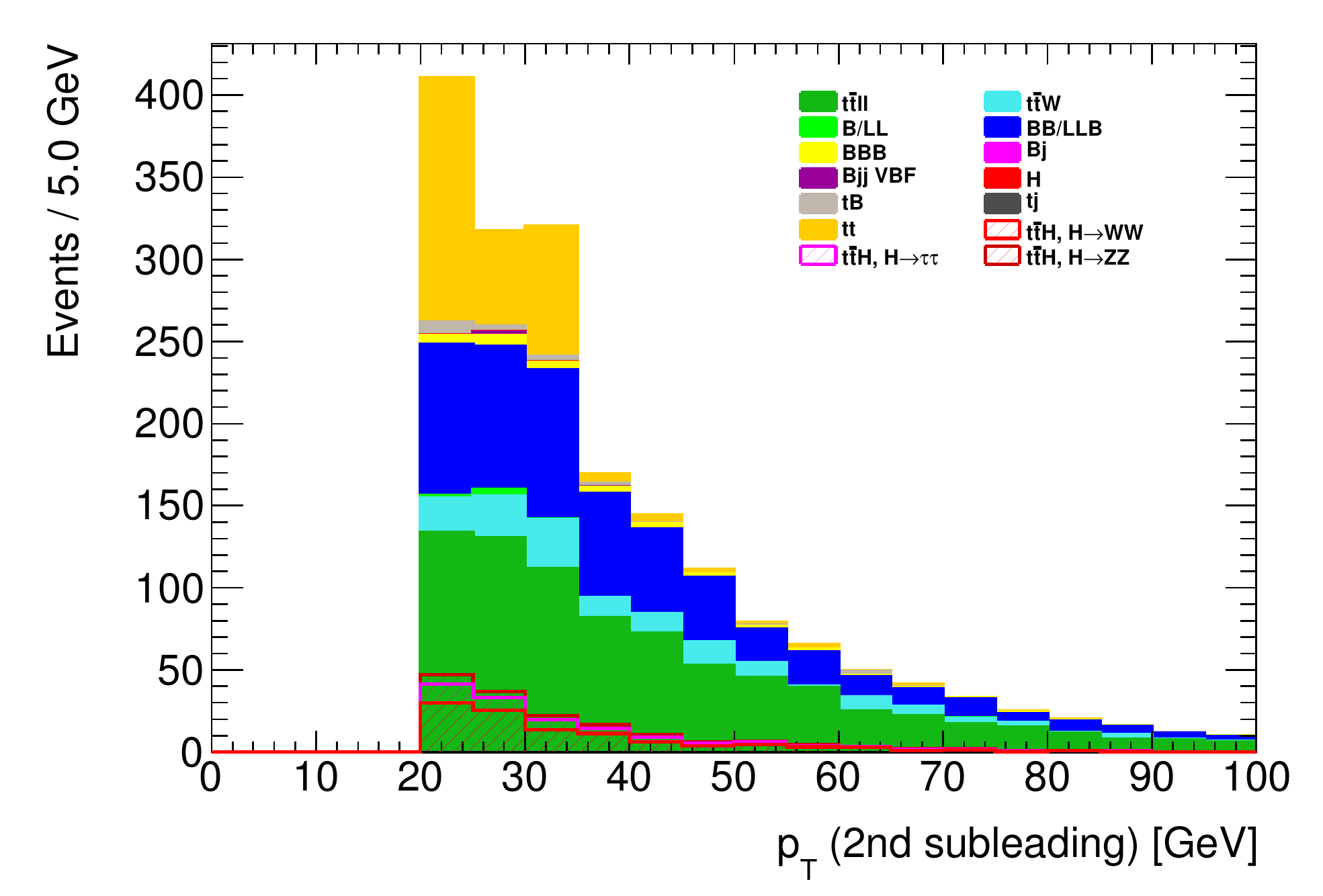}\\
\includegraphics[width=.5\linewidth]{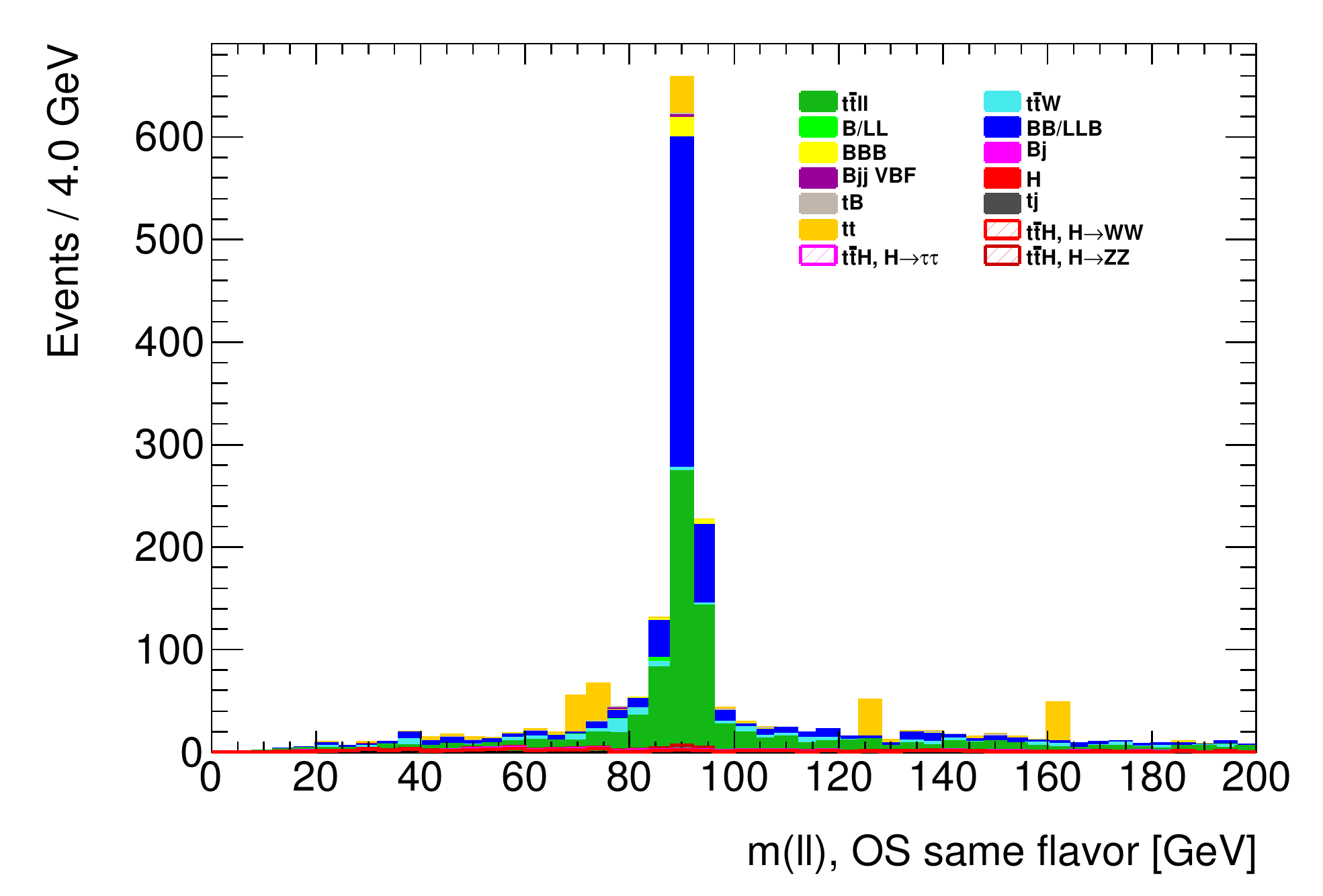}%
\includegraphics[width=.5\linewidth]{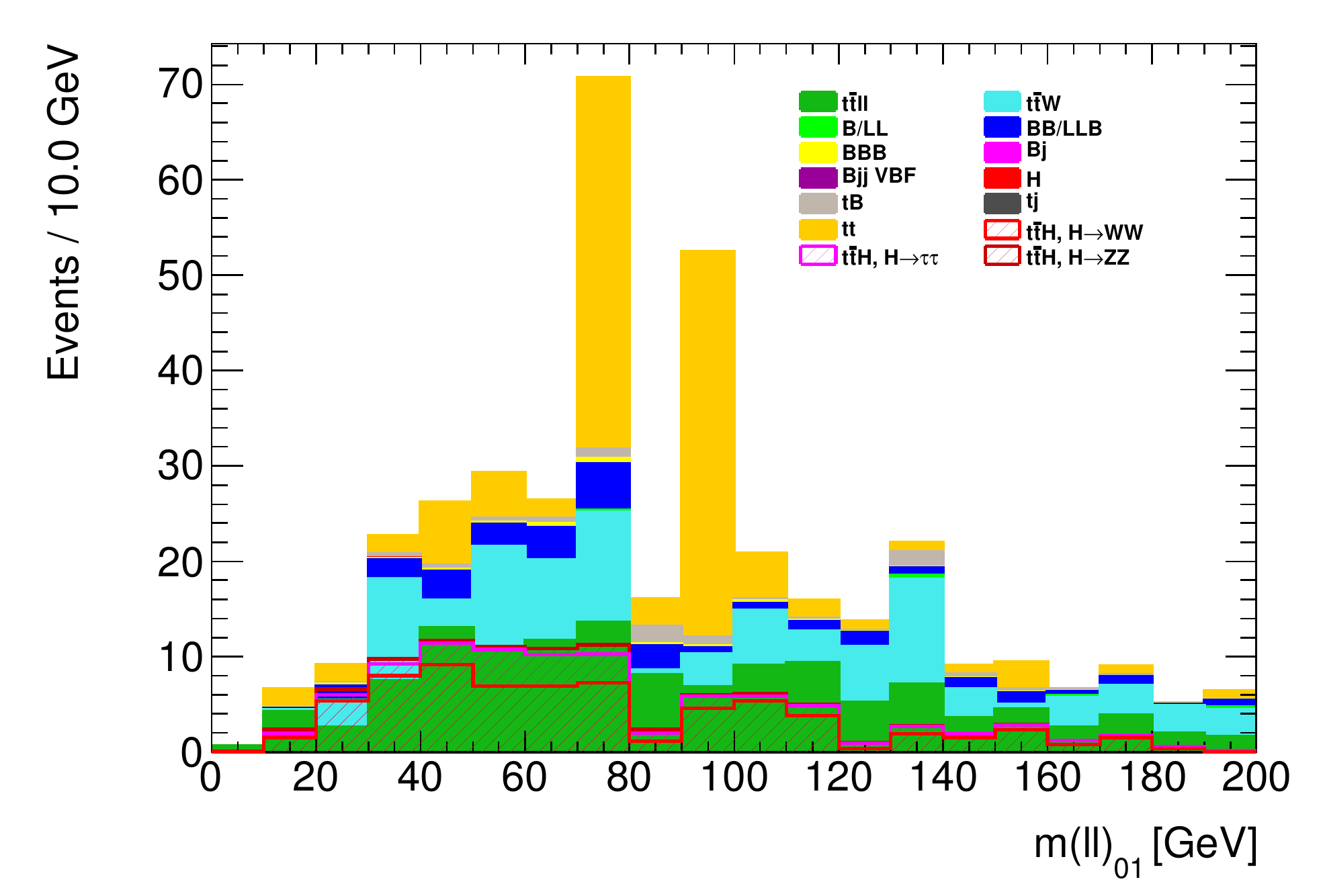}
\caption{\label{fig:3lplots}Selected plots from the 300 \ifb $3\ell$ analysis.  The top left plot shows the number of jets above 30 GeV after a loose $b$-tag is required.  The signal distribution is similar to that of $\ttbar Z$ and peaks at 4 jets.  Requiring 4 or more jets removes most diboson and \ttbar\ background.  The top right plot shows the \pt\ distribution of the lowest \pt\ lepton, after jet and btagging cuts.  The increase in \ttbar\ background at low \pt\ motivates the minimum \pt\ cut being raised from 20 to 25 GeV.  The bottom left plot shows the invariant mass distribution of all pairs of opposite sign, same flavor leptons.  Events can contribute 0, 1, or 2 times to this plot.  A large amount of diboson and $\ttbar Z$ background can be rejected by vetoing a region around the $Z$ mass.  The bottom right plot shows the distribution of the $m(\ell\ell)_{01}$ variable (the invariant mass of the lepton pair with smallest $\Delta R$).  The signal peaks at lower values than the background and we separate the regions with $m(\ell\ell)_{01}$ less than and greater than 70 GeV for fitting.  Note that the \ttbar\ prediction is subject to large MC statistics fluctuations and was not used to choose the 70 GeV boundary.}
\end{center}

\end{figure}

\begin{figure}
\begin{center}
\includegraphics[width=.5\linewidth]{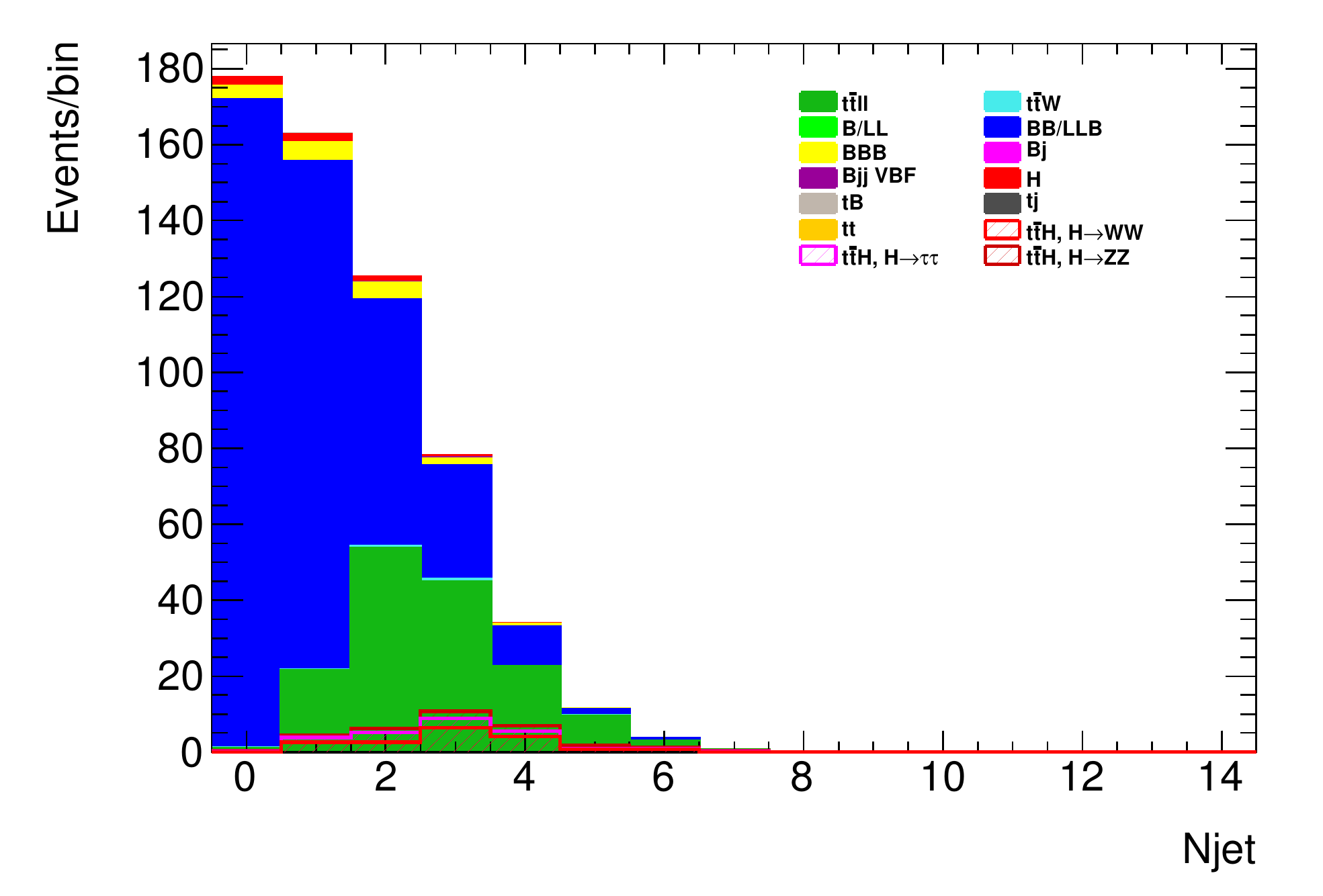}%
\includegraphics[width=.5\linewidth]{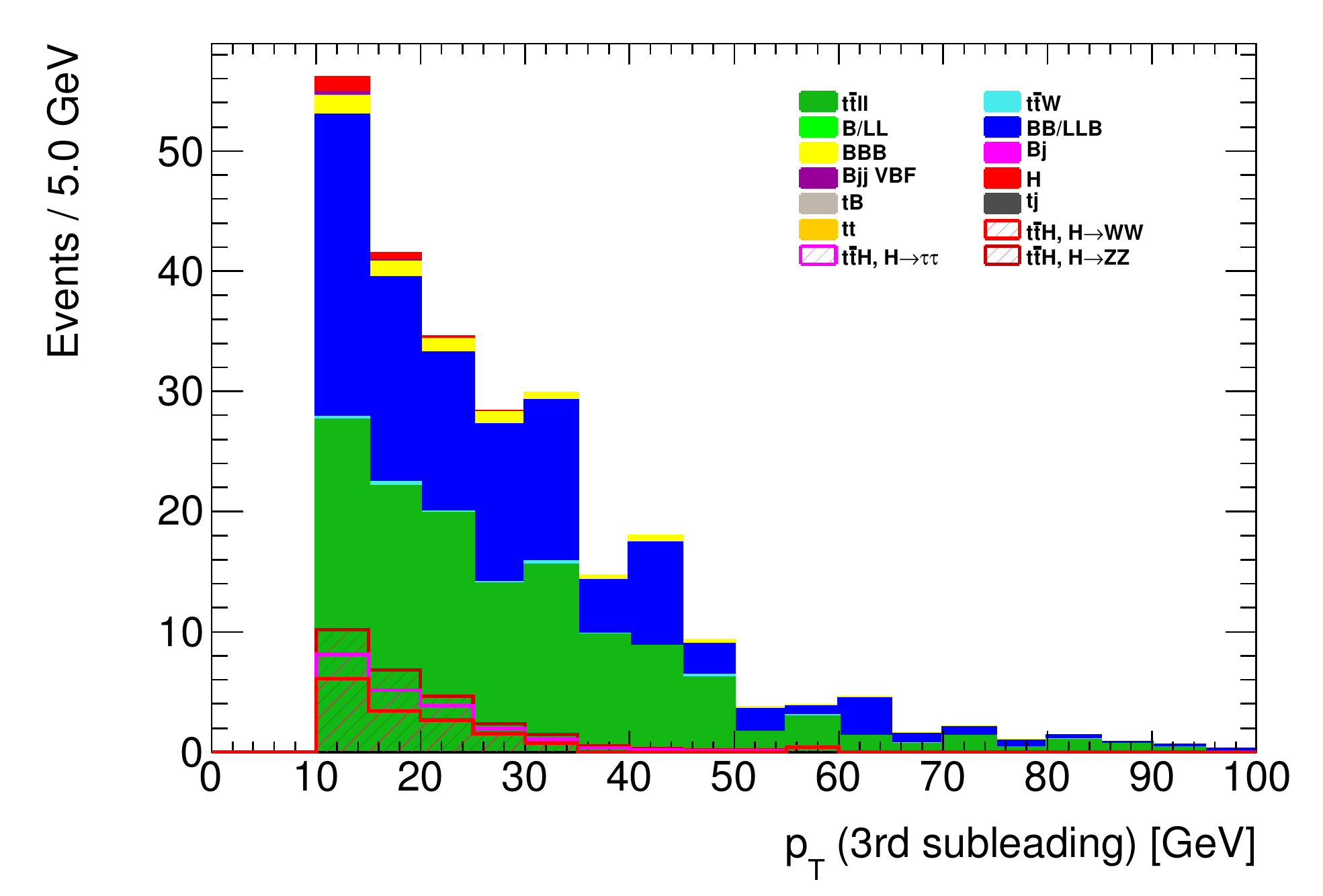}\\
\includegraphics[width=.5\linewidth]{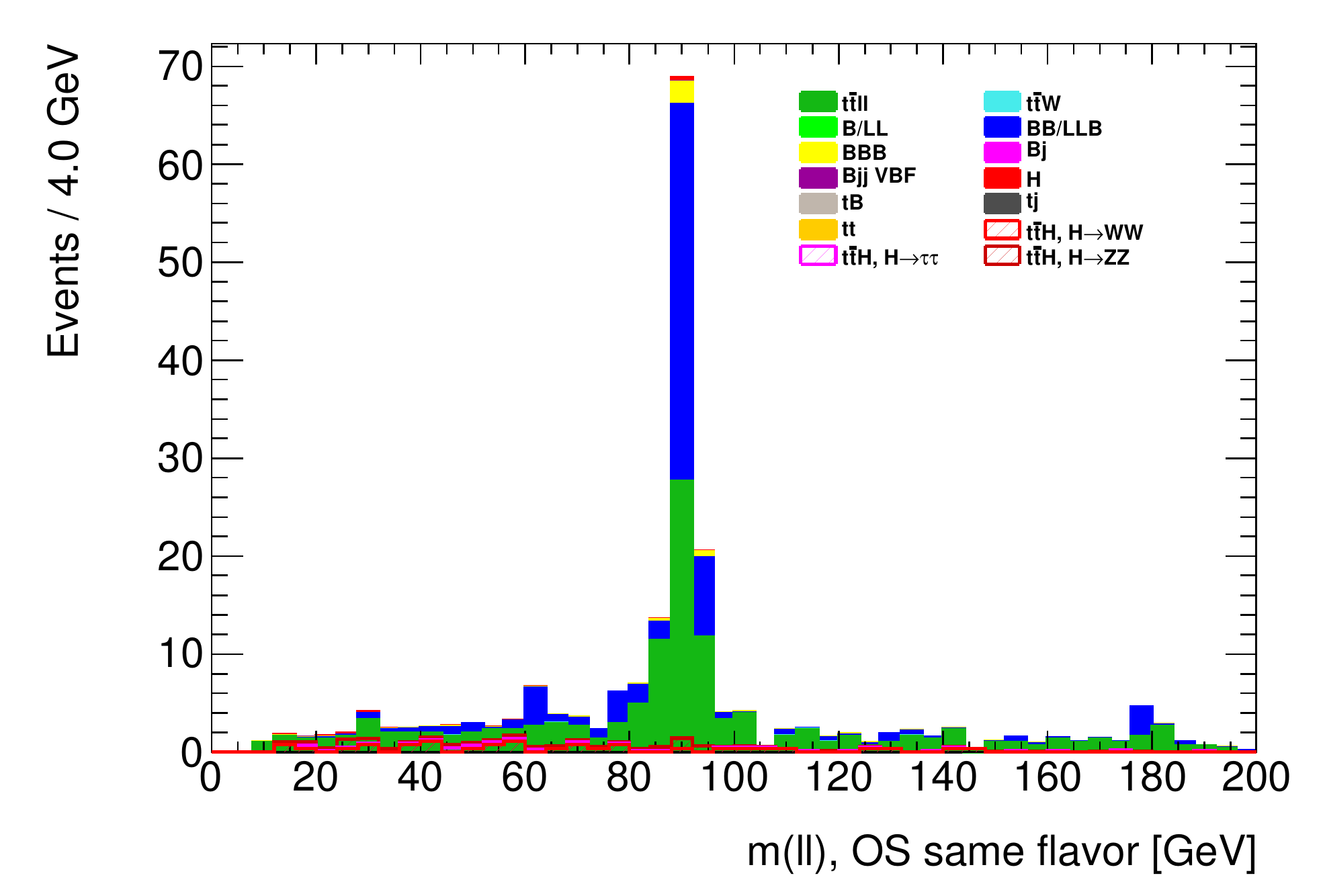}%
\caption{\label{fig:4lplots}Selected plots from the 300 \ifb $4\ell$ analysis.  The top left plot shows the number of jets above 35 GeV after a loose $b$-tag is required.  The signal distribution peaks at 3 jets.  We require 2 or more jets.  The top right plot shows the \pt\ distribution of the lowest \pt\ lepton, after jet and btagging cuts.  The importance of low-\pt\ acceptance is clear.  The bottom plot shows the invariant mass distribution of all pairs of opposite sign, same flavor leptons.  Events can contribute between 0 and 4 times to this plot.  A large amount of diboson and $\ttbar Z$ background can be rejected by vetoing a region around the $Z$ mass.}
\end{center}

\end{figure}

\begin{figure}
\begin{center}
 
 \includegraphics[width=.7\linewidth]{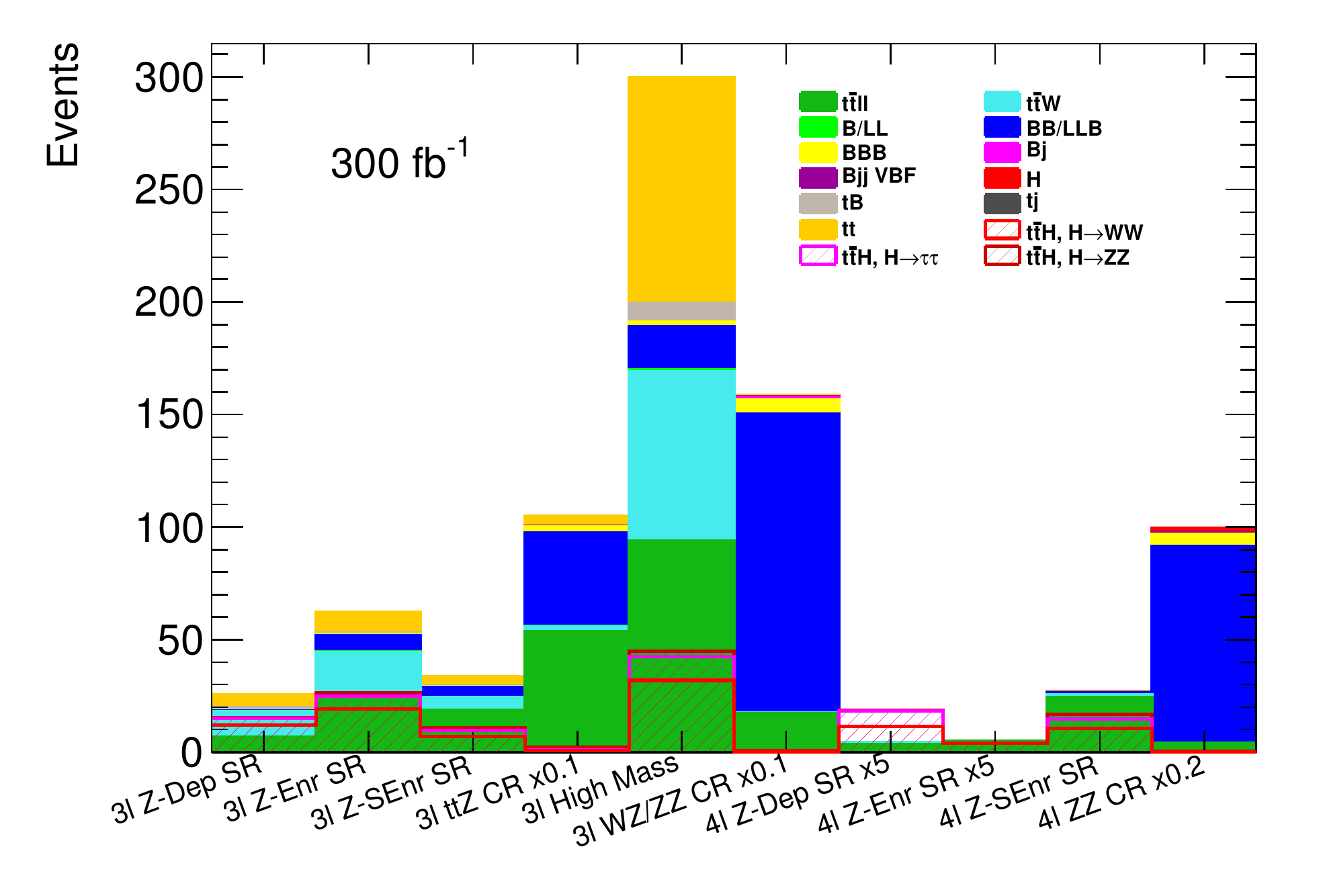}
\caption{\label{fig:yields}Yields of different processes in signal and control regions for 300 \ifb.  The definition of the regions is given in Table~\ref{tbl:cuts}.  Background processes are solid histograms and $\ttbar H$ signal processes are overlaid as hatched histograms.  Snowmass background processes are labeled as follows: B is an electroweak boson, LL is Drell-Yan dilepton production, j is a non-top parton, t is a top quark, H is a Higgs boson.}
\end{center}

\end{figure}

\begin{figure}
\begin{center}
 
 \includegraphics[width=.7\linewidth]{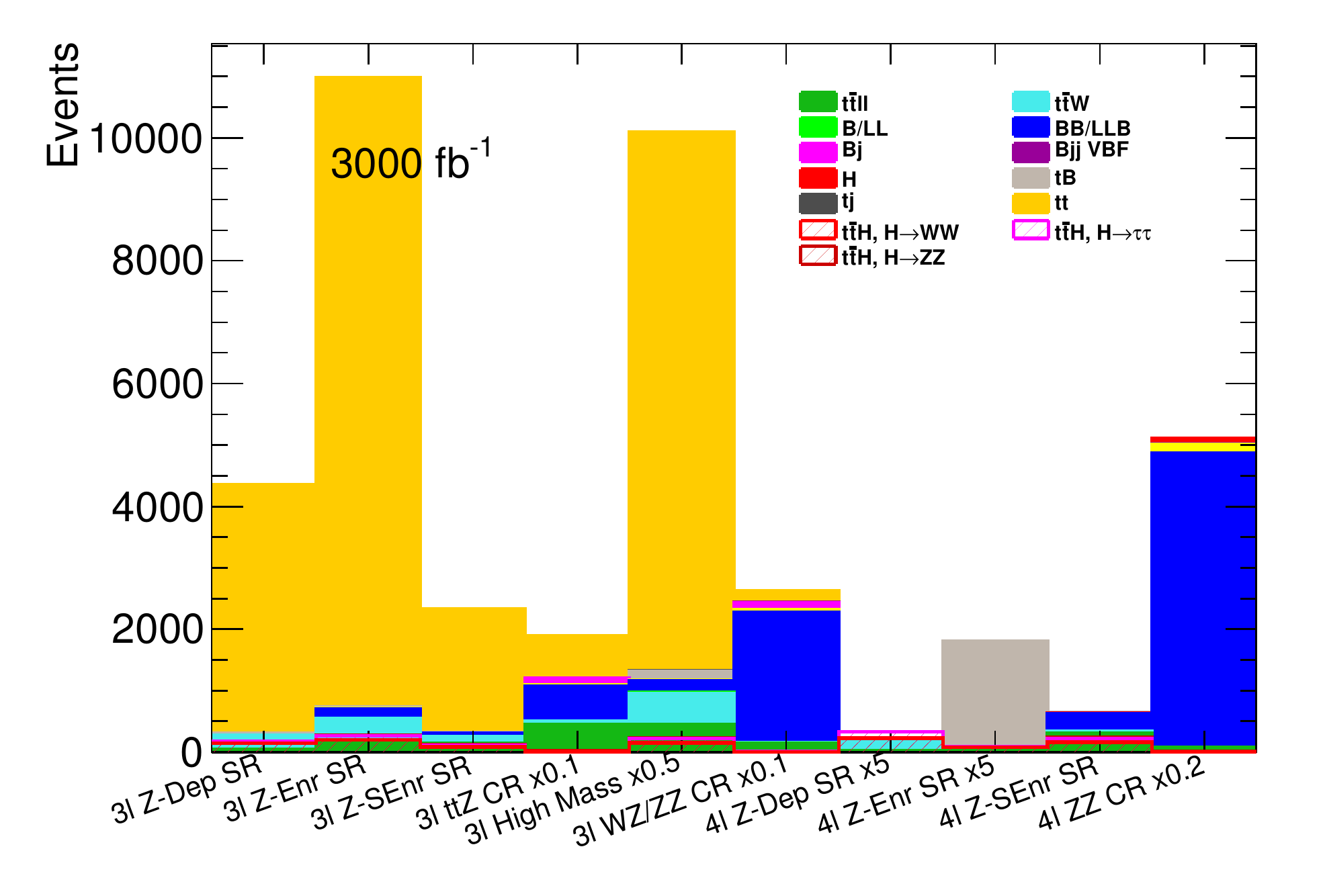}
\caption{\label{fig:3000yields}Yields of different processes in signal and control regions for 3000 \ifb.  The definition of the regions is given in Table~\ref{tbl:3000cuts}.  Background processes are solid histograms and $\ttbar H$ signal processes are overlaid as hatched histograms.  Snowmass background processes are labeled as follows: B is an electroweak boson, LL is Drell-Yan dilepton production, j is a non-top parton, t is a top quark, H is a Higgs boson.}
\end{center}

\end{figure}

We do not use the two same sign lepton signature, which has previously been put
forward as a sensitive channel 
\cite{Maltoni:2002jr,*Kostyukhin:2002txa,*Aad:2009wy,*Curtin:2013zua}.  The 
background levels in this channel depend strongly on the modeling of lepton fakes 
and of ``charge-flipped'' leptons which happen when the charge of one lepton is 
reconstructed incorrectly.  These are instrumental effects which are not well described by the 
currently available fast simulation.  We have confidence that this channel will prove important, but for this result we focus on the $3\ell$ and $4\ell$ channels where instrumental backgrounds are comparatively small.  For this reason the statistical sensitivity estimates are conservative.

As can be seen by comparing Figures~\ref{fig:yields} and \ref{fig:3000yields}, the higher pileup of the 3000 \ifb analysis introduces huge amounts of $t\bar t$ to the $3\ell$ analysis.  It is likely that significant improvements to the handling of jets and leptons in the presence of large pileup will be made by the time $\mu=140$ running occurs, in which case this estimate is unduly pessimistic.  With the current assumptions, the $3\ell$ channel is the main contributor to the 300 \ifb measurement, while the $4\ell$ channel dominates the 3000 \ifb result.

\section{Results}
 \label{sec:Results}
  \subsection{Statistical Method}
\label{sec:stat}

We use the HistFactory component of RooStats \cite{Cranmer:2012sba} to construct a likelihood function for the total yields in the ten signal/control regions.  Within each region we count events.  In other words, we do not use the distinctions 
between signal and background differential distributions in a fit to extract 
the signal yield. 

We assume no uncertainty in the MC predictions for the yields when constructing the likelihood, as we assume that MC statistics will not be allowed to be a limiting systematic uncertainty.  In all fits (including statistical-only) we 
place a loose prior on the $\ttbar Z$ ($\pm 10.7$\%, see Table~\ref{tab:syst}) 
and $WZ/ZZ$ ($\pm 30$\%) normalization, which are then constrained 
automatically during the fit by the control regions.

The estimated precision on the parameter $\mu(\ttbar H) \equiv \sigma_\textrm{best fit}/\sigma_\textrm{SM}$ is obtained by performing 1000 pseudoexperiments, fluctuating all nuisance parameters and the total event yield and fitting the dataset, and computing the root-mean-square (RMS) of the obtained values of $\mu(\ttbar H)$.  The pulls reported by the fits are checked and have RMS compatible with 1, indicating proper statistical behavior of the fits.

The \ttbar\ samples have large fluctuations due to single events having large weights.  This can lead the fit to believe that there are significant differences in the distribution of signal and \ttbar\ between the various signal regions and to use this information to constrain the \ttbar\ fake rate.  To avoid this behavior, for the fit we assume that the \ttbar\ contribution in each signal or control region is proportional to the Higgs signal contribution, with total yield equal to that of the total expected \ttbar\ yield.  This procedure makes the \ttbar\ impossible to distinguish from signal via the fit within each channel, and provides a conservative estimate of the effect of \ttbar\ lepton fake rate uncertainties.

\begin{table}
\begin{center}
\caption{\label{tbl:statonly} Statistical uncertainties for $3\ell$, $4\ell$, and combined measurements of $\mu(\tth)$, projected for 300 and 3000 \ifb.}
 \begin{tabular}{lcc}
 \hline\hline
  {\sc Channel} & 300 \ifb & 3000 \ifb\\
  \hline
  $3\ell$ only & 25\% & ---\% \\
  $4\ell$ only & 34\% & 12\% \\
  Combined & 21\% & 9\% \\
  \hline\hline
 \end{tabular}
\end{center}
\end{table}

\subsection{Systematic Uncertainties}
\label{sec:syst}

It is important that NLO calculations be available for this analysis
since the LO theoretical uncertainties are very large.  For instance,
the scale uncertainty for the \ttbar$+\ell\ell$ process is $^{+57}_{-24}\%$ at leading order.
We provide NLO scale uncertainties for signal and physics background
processes in Table~\ref{tab:syst}. 
We symmetrize the \ttbar$W$ uncertainty and use this also for the 
\ttbar$+\ell\ell$ because of the similar mass scales of the produced system.
The NLO calculation of \ttbar$+\ell\ell$ scale uncertainty was 
computationally resource limited.  We also incorporate a signal uncertainty
of $\pm 8.9\%$, taken from Ref.~\cite{Dittmaier:2011ti}, added linearly to the scale uncertainty.

\begin{table}
\begin{center}
\caption{Next-to-leading order scale uncertainties for signal and two main
physics backgrounds.  aMC@NLO was used for these calculations.} 
\label{tab:syst}
\begin{tabular}{ccccc}
\hline\hline
{\sc Process} & {\sc NLO uncert} \\
\hline
\ttbar $H$    & $+5.4\%-8.8\%$ \\
\ttbar $W$    & $+10.9\% -10.4\%$ \\
\ttbar $ll$   & $\pm10.7\%^*$ \\
\hline\hline
\end{tabular}
\end{center}
\end{table}

We assume a 30\% relative uncertainty on the \ttbar\ fake rate.  The \ttbar\ contribution is significant in the $3\ell$ analysis, and it is here that the cleaner $4\ell$ channel can contribute to reduce the systematic uncertainty in the combination.    We find that the fake rate is still constrained by the data, because \ttbar\ and $\ttbar H$ contribute differently to the $3\ell$ and $4\ell$ channels and the fit is able to exploit this difference while improving the uncertainty on both.  We feel that this is a reasonable constraint based on the underlying physics and so do not try to avoid this behavior in the fit.

\begin{table}
 \begin{center}
\caption{\label{tbl:syst2} Systematic uncertainties on combined measurement of $\mu(\tth)$, projected for 300 and 3000 \ifb. Due to correlations, the combined systematics are not exactly the quadrature sum of individual components.}
  \begin{tabular}{lcc}
   \hline\hline
  {\sc Source} & 300 \ifb & 3000 \ifb \\
  \hline
  Top fake rate & 17\% & 2\% \\
  $\sigma(\ttbar H)_\textrm{SM}$ & 16\% & 16\%\\
  Other cross section systematics & 8\% & 3\%\\
  \hline
  All systematics & 27\% & 17\% \\
  Systematics without $\sigma(\ttbar H)_\textrm{SM}$ & 18\% & 4\% \\
  \hline\hline
 \end{tabular}
 \end{center}
\end{table}

\subsection{Projected Sensitivity in 300 and 3000 fb$^{-1}$}

The statistics-only predicted precision for 300 and 3000 \ifb is shown in 
Table~\ref{tbl:statonly}, and the impact of various systematic effects is shown 
in Table~\ref{tbl:syst2}.  It can be seen that dramatic improvement in the Higgs cross section prediction (presumably by going to NNLO) is necessary in order to match the achievable statistical uncertainties with 3000 \ifb.  The effort to implement these improvements will be important for the 
measurement described in this paper.
Other cross section uncertainties are constrained by data and improve with luminosity. The $t\bar t$ fake rate becomes a minor contribution once the fit is dominated by the $4\ell$ channel, which has very little $t\bar t$ background even for $\langle\mu\rangle=140$.

Ignoring the $\ttbar H$ cross section uncertainty, we predict an overall precision (combining statistical and systematic uncertainties in quadrature) of 28\% (10\%) after 300 \ifb (3000 \ifb).  We assume that we can take $\mu(\ttbar H) \propto Y_t^2$, so the uncertainty on $Y_t$ $\approx$ half the uncertainty on $\mu(\ttbar H)$.  The cross section measurement precisions above thus correspond to a Yukawa coupling precision of 14\% (5\%).  Our current knowledge of $\sigma(\ttbar H)_\textrm{SM}$ adds $\sim 8\%$ in quadrature to this.

We have not yet accounted for detector systematics, in particular those arising from jet counting, $b$-tagging, or lepton efficiencies in the high pileup environment of HL-LHC.  It is 
anticipated that enough information will be available at that point to make those systematics small compared to the cross section uncertainties.  In addition it should be noted that this analysis is not fully optimized (in particular with respect to fake lepton rejection). Analysis improvements may make the ultimate precision better than what has been described.
\label{sec:proj}

\section{Conclusion}
 \label{sec:Conclu}
  We have studied the sensitivity of the LHC to extract the Yukawa coupling
of the top quark from \ttbar$H$ events, reconstructed in the $3\ell$ and 
$4\ell$ channels.  We find that the analysis has high statistical power 
which may start to become limited by systematic uncertainties even with
as little luminosity as 300 \ifb.  The $\ttbar +$ jets background poses a significant challenge for exploiting the $3\ell$ channel with the HL-LHC, where the sensitivity is dominated by the $4\ell$ channel.  Improvement in the prediction of 
$\sigma(\ttbar H)_\textrm{SM}$ is definitely required to take full advantage 
of HL-LHC statistics.  With current uncertainties on the $\ttbar H$ 
cross section, we expect 
ultimately for each LHC detector to achieve a precision on the top 
Yukawa coupling of approximately 9\% in the $3\ell+4\ell$ analysis alone without improvement on the uncertainty on $\sigma(\ttbar H)$, or 6\% if this is improved by a factor of two.

\bibliographystyle{atlasBibStyleWoTitle}
\bibliography{tth}

\end{document}